\documentstyle[12pt]{article}
\newcommand\mysection{\setcounter{equation}{0}\section}

\def\a0{\bar\alpha_0}
\def\ae{\alpha_{\mbox{\scriptsize eff}}}

\def\as{\alpha_{\mbox{\tiny S}}}
\def\b0{\beta_0}

\def\cd{\chi^2/\mbox{d.o.f.}}
\def\ee{e^+e^-}
\def\enap{\mbox{e}}

\def\GeV{\mbox{\rm{GeV}}}
\def\kp{k_\perp}
\def\kps{k_\perp^2}

\def\lms{\Lambda^{(5)}_{\overline{\mbox{\tiny MS}}}}
\def\mI{\mu_{\mbox{\tiny I}}}
\def\mR{\mu_{\mbox{\tiny R}}}
\def\MSbar{\overline{\mbox{MS}}}

\def\pt{{\mbox{\scriptsize pert}}}

\def\st{\sigma_{\mbox{\scriptsize tot}}}

\def\cO#1{{\cal{O}}\left(#1\right)}
\def\frac#1#2{ {{#1} \over {#2} }}

\def\VEV#1{\left\langle #1\right\rangle}
\def\beq{\begin{equation}}
\def\beeq{\begin{eqnarray}}
\def\eeq{\end{equation}}
\def\eeeq{\end{eqnarray}}
\newskip\humongous \humongous=0pt plus 1000pt minus 1000pt
\def\caja{\mathsurround=0pt}
\def\eqalign#1{\,\vcenter{\openup1\jot
\caja   \ialign{\strut \hfil$\displaystyle{##}$&$
\displaystyle{{}##}$\hfil\crcr#1\crcr}}\,}

\newif\ifdtup

\def\cpc#1#2#3{Comp.\ Phys.\ Commun.\ #1 (19#3) #2}

\def\np#1#2#3{Nucl.\ Phys.\ B#1 (19#3) #2}
\def\pl#1#2#3{Phys.\ Lett.\ #1B (19#3) #2}
\def\pr#1#2#3{Phys.\ Rev.\ D#1 (19#3) #2}
\def\prl#1#2#3{Phys.\ Rev.\ Lett.\ #1 (19#3) #2}
\def\zp#1#2#3{Z.\ Phys.\ C#1 (19#3) #2}
\begin{document}
\begin{titlepage}
\renewcommand{\thefootnote}{\fnsymbol{footnote}}
\begin{flushright}
     Cavendish--HEP--97/2 \\
     hep-ph/9704298
\end{flushright}
\vspace*{\fill}
\begin{center}
{\Large \bf
Power Corrections to\\[1ex]
Event Shape Distributions\footnote{Research
supported in part by the U.K.\ Particle Physics and
Astronomy Research Council and by the EC Programme
``Training and Mobility of Researchers", Network
``Hadronic Physics with High Energy Electromagnetic Probes",
contract ERB FMRX-CT96-0008.}}
\end{center}
\par \vskip 2mm
\begin{center}
        {\bf Yu.L.\ Dokshitzer\footnote{Permanent address:
St Petersburg Nuclear Physics Institute, Gatchina, St Petersburg 188350,
Russian Federation.}} \\
         INFN Sezione di Milano, Via Celoria 16, 20133 Milan, Italy\\[1.5ex]
        {\bf B.R.\ Webber} \\
        Cavendish Laboratory, University of Cambridge,\\
        Madingley Road, Cambridge CB3 0HE, U.K.
\end{center}
\par \vskip 2mm
\begin{center} {\large \bf Abstract} \end{center}
\begin{quote}
We estimate the effects of non-perturbative physics
on the differential distributions of infrared- and
collinear-safe $\ee$ event shape variables, by
extending the notion of an infrared-regular effective
strong coupling, which accounts for the non-perturbative
corrections to the mean values of several shape
variables, to their distributions. This leads to $1/Q$
power corrections over a range of values of the shape
variables considered, where $Q$ is the centre-of-mass energy.
In the case of the thrust variable, the leading correction is
simply a shift of the distribution, by an amount proportional
to $1/Q$. We show that this gives an excellent description
of the data throughout a wide range of $T$ and $Q$.
\end{quote}
\vspace*{\fill}
\begin{flushleft}
     Cavendish--HEP--97/2\\
     April 1997
\end{flushleft}
\end{titlepage}
\renewcommand{\thefootnote}{\fnsymbol{footnote}}

\mysection{Introduction}
The study of power-suppressed corrections to QCD observables
has become a lively field of experimental and theoretical
investigation. On the experimental side, the estimation of
power-suppressed hadronization or higher-twist corrections
is necessary for the accurate measurement of the strong
coupling $\as$, a crucial parameter of the Standard Model.
Theoretical work on divergences of the QCD
perturbation series, in particular infrared renormalons
[\ref{renormalons},\ref{BBB}], and related attempts to define
the running coupling beyond perturbation theory
[\ref{effcharge}], have also led to renewed
interest in power-suppressed contributions.

Power corrections to hadronic event shapes are of particular
interest, because they are expected to have a characteristic
$1/Q$-dependence on the hard process scale $Q$ (the centre-of-mass
energy in $\ee$ annihilation) [\ref{Web94}-\ref{BraBen}].
Thus, in contrast to the $1/Q^2$ corrections encountered
in most quantities, they are still important even at
$Q\sim M_Z$, where most of the $\ee$ data have been obtained.
The theory of $1/Q$-corrections is much less developed than
that for higher powers; in particular, the operator product
expansion does not apply in this case, since there are no
relevant operators of the corresponding dimension.

In refs.~[\ref{Web94},\ref{DokWeb95},\ref{BPY}] a treatment of
$1/Q$-corrections to event shapes was proposed, based on the
notion of an effective strong coupling, $\ae$, which is
approximately universal but differs from the perturbative
form in the infrared region (see also [\ref{DKT}]).
Such an approach was found to be 
quite successful in describing the powers and approximate
magnitudes of power corrections to a wide variety
of QCD observables, using the low-energy moments of $\ae$
as non-perturbative parameters.  The comparison with data on
event shapes, however, has been limited so far to the mean values
only [\ref{DokWeb95},\ref{shapexp}].

In the present paper we apply the same approach to the differential
distributions of event shape variables. The central idea remains
that of ref.~[\ref{DokWeb95}]: the emission of soft gluons is
assumed to be controlled by an effective coupling $\ae$,
different from the perturbative form in the infrared region
but small enough for terms of higher order in
$\ae$ to be neglected as a first approximation. We combine
this idea with the treatment of event shape distributions
developed in refs.~[\ref{thrust},\ref{CTTW}].  It was
shown there that some shape variables have the property of
exponentiation, which allows large logarithms to be resummed
to all orders in perturbation theory. For such variables,
the effective coupling assumption means that the leading
non-perturbative corrections also exponentiate, implying
a specific transformation of the whole distribution, or
at least its logarithmically enhanced part. This has been
pointed out in ref.~[\ref{KorSte}].

We concentrate here, as in ref.~[\ref{DokWeb95}], on the thrust
variable, $T$ [\ref{tdef}]. In this case the leading non-perturbative
effect over a range of thrust values turns out to be simply a shift in
the distribution, by an amount proportional to $1/Q$,
modulo logarithmic $Q$-dependence. The shift is just
such that we recover the result of ref.~[\ref{DokWeb95}]
for the mean value.  Remarkably, this leads
to an excellent description of the data
over a wide range of $T$ and $Q$.  The only two free
parameters are the non-perturbative quantity
\beq
\a0(\mI)\equiv\frac{1}{\mI}\int_0^{\mI}dq\,\ae(q)
\eeq
which characterises the behaviour of the effective coupling
below some infrared matching scale $\mI$, and the
perturbative coupling $\as(M_Z)$.  For these quantities we
find values consistent with those obtained from other data.

After a detailed consideration of the thrust distribution,
we discuss the relation between our analysis and the
``dispersive approach" of ref.~[\ref{BPY}], and then
comment briefly on the prospects for extending the method to
other shape variable distributions.

\mysection{Thrust distribution}
It was shown in ref.~[\ref{CTTW}] that the thrust distribution
is given to next-to-leading logarithmic accuracy by the expression
\beq\label{dsdT}
\frac{1}{\st}\frac{d\sigma}{dT} \equiv F(T) =
\frac{Q^2}{2\pi i}\int_C d\nu\,\enap^{(1-T)\nu Q^2}
\left[\tilde J_\nu^q(Q^2)\right]^2\;,
\eeq
where the contour $C$ runs parallel to the imaginary axis,
to the right of all singularities of the integrand,
and $\tilde J_\nu^q(Q^2)$ represents the
Laplace transform of the quark jet mass distribution
at hard process scale $Q$.  The result obtained
for this function, again to next-to-leading logarithmic
accuracy, was
\beq\label{lnJ}
\ln\tilde J_\nu^q(Q^2) = \int_0^1\frac{du}{u}
\left(\enap^{-u\nu Q^2}-1\right)
\left[\int_{u^2Q^2}^{uQ^2}\frac{dq^2}{q^2}
A(\as(q))+\frac 1 2 B(\as(\sqrt{u}Q))\right]
\eeq
with
\beq
A(\as) = C_F\frac{\as}{\pi}\left(1+K\frac{\as}{2\pi}\right)
\;,\;\;\;
B(\as) = -3C_F\frac{\as}{2\pi}\;,\;\;\;
K= C_A\left(\frac{67}{18}-\frac{\pi^2}{6}\right) - \frac{5}{9}N_f\;,
\eeq
and $\as$ defined in the $\MSbar$ scheme. 

In ref.~[\ref{CTTW}], the region of low $q^2$ in Eq.~(\ref{lnJ})
was ignored, on the grounds that its contribution is subleading.
We now include it in the following way: we first subtract the
perturbative contribution from the region $q^2<\mI^2$, and
then add it back again using the non-perturbative effective
coupling $\ae(q)$.  Changing the order of integration, the
amount added back is
\beq\label{dlnJexp}
\delta\ln\tilde J_\nu^q(Q^2) =\frac{2 C_F}{\pi}
\int_0^{\mI}\frac{dq}{q}\ae(q)\int_{q^2/Q^2}^{q/Q}
\frac{du}{u}\left(\enap^{-u\nu Q^2}-1\right)\;.
\eeq
Since $\nu Q^2$ is conjugate to $1-T$, for $1-T\gg\mI/Q$ we
can safely expand the exponential to first order, to obtain
\beq\label{dlnJlin}
\delta\ln\tilde J_\nu^q(Q^2) \simeq -\frac{2 C_F}{\pi}
\int_0^{\mI}dq\,\ae(q)\,\nu Q
\equiv -\frac{2 C_F}{\pi}
\frac{\mI}{Q}\,\a0(\mI)\,\nu Q^2 \;.
\eeq
The term involving $B(\as)$ gives a correction
of order $1/Q^2$, which we neglect, together with terms
of order $\ae^2$. 

For the perturbative subtraction, we use the next-to-leading-order
expansion of $\as(q)$ in terms of $\as(\mR)$, $\mR$ being the
chosen renormalization scale:
\beq
\as(q) = \as(\mR) + \frac{\b0}{2\pi}\ln\frac{\mR}{q}\,\as^2(\mR) 
\eeq
where $\b0 = (11 C_A-2N_f)/3$. The expression to be
subtracted from the right-hand side of Eq.~(\ref{dlnJlin})
is thus
\beq
-\frac{2 C_F}{\pi}\int_0^{\mI}dq
\left[\as(\mR)+\frac{\b0}{2\pi}\left(\ln\frac{\mR}{q}
+\frac{K}{\b0}\right)\as^2(\mR)\right]\nu Q\;,
\eeq
which gives
\beq
\delta\ln\tilde J_\nu^q(Q^2)=-\frac{2 C_F}{\pi}
\frac{\mI}{Q}\left[\a0(\mI)-\as(\mR)-\frac{\b0}{2\pi}
\left(\ln\frac{\mR}{\mI}+\frac{K}{\b0}+1\right)\as^2(\mR)\right]
\nu Q^2 \;.
\eeq

Substituting in Eq.~(\ref{dsdT}), we see that the leading
non-perturbative effect on the thrust distribution is simply
to shift the perturbative prediction to lower thrust,
by an amount proportional to $1/Q$:
\beq\label{Tshift}
F(T) = F^\pt(T-\delta T)
\eeq
where
\beq\label{deltaT}
\delta T = -\frac{4 C_F}{\pi}
\frac{\mI}{Q}\left[\a0(\mI)-\as(\mR)-\frac{\b0}{2\pi}
\left(\ln\frac{\mR}{\mI}+\frac{K}{\b0}+1\right)\as^2(\mR)\right]\;.
\eeq
This is precisely the formula derived in ref.~[\ref{DokWeb95}]
for the non-perturbative shift in the mean thrust, $\delta\VEV{T}$.

We remind the reader that the simple prediction (\ref{Tshift})
applies only in the region $1-T\gg\mI/Q$, where $\mI$ marks the
scale below which $\ae$ starts to deviate from $\as$. For
$1-T\sim\mI/Q$ one could explore the effects of retaining
more terms in the expansion of the exponential function in
Eq.~(\ref{dlnJexp}), but this would require a detailed
parametrization of $\ae$, and we have not tried it.
In practice, we excluded the region $1-T<0.05$ from all
our comparisons with data.

The prediction (\ref{Tshift}) also does not strictly apply at
low values of the thrust, where terms involving powers
of $\ln(1-T)$ are not dominant. In ref.~[\ref{CTTW}],
however, it was found that exponentiation has significant
effects throughout the region in which the cross section
is substantial. In earlier comparisons with data, good
agreement was obtained (after hadronization corrections)
by using a `$\log R$' matching scheme, in which essentially
all known higher-order corrections are exponentiated.
It therefore appears natural to adopt this scheme and to
extend our comparisons with Eq.~(\ref{Tshift}) rather
far into the low-thrust region. In fact we find good
agreement down to $T=0.65$, which is even outside the
three-jet region ($T>2/3$).

For $F^\pt$ we take the $\log R$-matched resummed expression in
ref.~[\ref{CTTW}] with no modification of the logarithmic terms,
i.e.\ with $L= -\ln(1-T)$. Initially, we set the renormalization
scale $\mR = Q$ and the infrared matching scale $\mR = 2$ GeV, as in
ref.~[\ref{DokWeb95}]. The resulting predictions were compared
with data on the thrust distribution in the interval $0.05<1-T<0.35$
at energies $14<Q<161$ GeV, as listed in Table~1.

\begin{figure}[htb]
\vspace{12.5cm}
\includegraphics{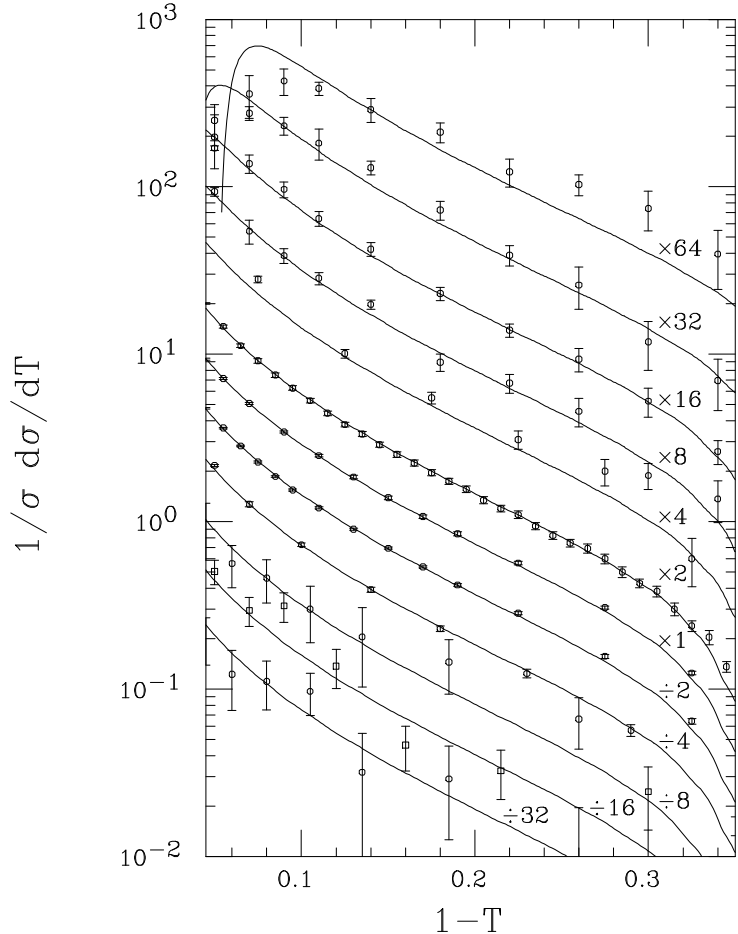}
\caption{Data on the thrust distribution at various energies, and the
prediction (\ref{Tshift}) for $\lms = 0.235$ GeV, $\a0 = 0.46$.
The twelve data sets are as listed in Table 1 (from top to bottom,
multiplied by the factors indicated).}\label{tfits}
\end{figure}

\begin{figure}[htb]
\vspace{11.5cm}
\includegraphics{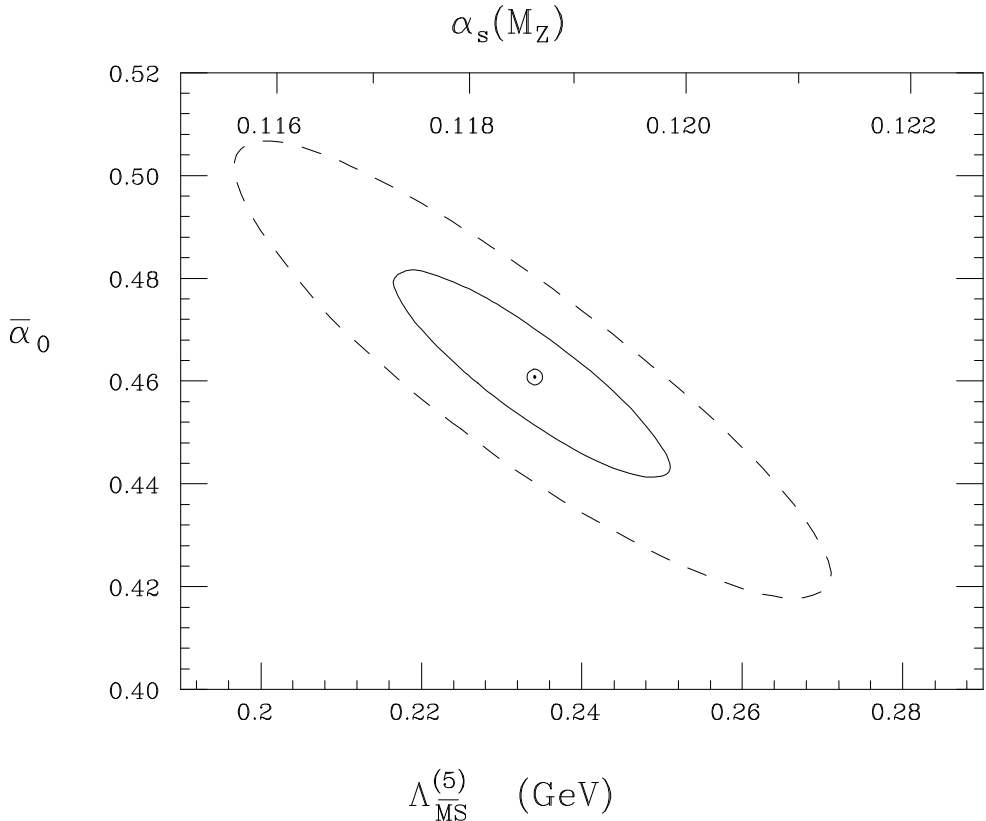}
\caption{Best fit values and 95\% confidence region for the two fitted
parameters. Solid/dashed ellipses: including/excluding 14 GeV data.}
\label{ellipse}
\end{figure}

The best fit values of the two free parameters are
\beq
\lms = 0.235 \pm 0.017\ \GeV\;,\;\;\;\a0(2\ \GeV) = 0.46 \pm 0.02
\eeq
(95\% confidence level). The corresponding curves are shown in
Fig.~\ref{tfits}.  The fitted values of the parameters are somewhat
correlated, as shown in Fig.~\ref{ellipse}.
The value of $\lms$, corresponding to $\as(M_Z) = 0.1186 \pm 0.0013$,
is in good agreement with that obtained by other methods. The value
of $\a0$ is somewhat smaller than, but within two standard deviations
of, that obtained in ref.~[\ref{DokWeb95}],
$\a0(2\ \GeV) = 0.52\pm 0.03$.

The quality of the overall fit is remarkable ($\cd = 134/114 = 1.18$)
-- much better than those typically obtained\footnote{Compare,
for example, the best fit $\cd = 2.3$ obtained in
ref.~[\ref{tdat3}] at a single energy for the more restricted
interval $0.06<1-T<0.30$.} when hadronization
effects are estimated from Monte Carlo models [\ref{Jetset},\ref{Herwig}].
Furthermore, we see from Table~1 that a large contribution to the $\chi^2$
comes from the data at 14 GeV, where our fitting region is perhaps too
large ($\mI/Q = 0.14>0.05$) and there may be complications due to quark
mass effects, heavy quark decays and higher power corrections.
However, the 14 GeV data are valuable because they provide the
longest lever arm for distinguishing between inverse power and
logarithmic energy dependence. If we exclude the 14 GeV data
altogether, the best fit parameter values do not change, but
the errors are doubled (dashed curve in Fig.~\ref{ellipse}).

To study the dependence on the infrared matching scale, we also
performed a fit at $\mI=3$ GeV. The best fit value of $\lms$ and
the quality of the fit did not change significantly, and we
obtained $\a0(3\ \GeV) = 0.374 \pm 0.010$, again within two
standard deviations of the value obtained in ref.~[\ref{DokWeb95}],
$\a0(3\ \GeV) = 0.42\pm 0.03$.

We also investigated the dependence on the the renormalization
scale $\mR$, in the range $Q^2/2<\mR^2<2Q^2$. The best fit parameter
values varied from $\lms = 0.204$ GeV, $\a0(2\ \GeV) = 0.457$ to
$\lms = 0.270$ GeV, $\a0(2\ \GeV) = 0.466$, respectively.
Thus the error in $\lms$ is still dominated by the systematic error
due to renormalization scale dependence, and our overall estimate
of this parameter is
\beq
\lms = 0.235 \pm 0.035\ \GeV\;,\;\;\;\as(M_Z) = 0.1185 \pm 0.0025\;.
\eeq

\begin{table}[htb]
\caption{Data sets and fit results for
$\lms = 0.235$ GeV, $\a0 = 0.46$.}
\begin{center}
\begin{tabular}{|c c c r r|}\hline
Collab. & $Q$/GeV  & Ref.       & Pts. & $\chi^2$\\ \hline
TASSO   & 14   & [\ref{tdat1}]  &  8   & 28.5 \\
TASSO   & 22   & [\ref{tdat1}]  &  8   & 11.1 \\
TASSO   & 35   & [\ref{tdat1}]  &  8   &  2.1 \\
TASSO   & 44   & [\ref{tdat1}]  &  8   &  7.6 \\
AMY     & 54   & [\ref{tdat2}]  &  6   & 12.5 \\
OPAL    & 91.2 & [\ref{tdat3}]  & 30   & 19.0 \\
ALEPH   & 91.2 & [\ref{tdat4}]  & 11   & 12.8 \\
DELPHI  & 91.2 & [\ref{tdat5}]  & 13   & 23.6 \\
SLD     & 91.2 & [\ref{tdat6}]  &  6   &  3.5 \\
OPAL    & 133  & [\ref{tdat7}]  &  6   &  2.9 \\
DELPHI  & 133  & [\ref{tdat8}]  &  6   &  8.4 \\
OPAL    & 161  & [\ref{tdat9}]  &  6   &  2.3 \\ \hline
Total   &      &                &116   &134.1 \\
\hline \end{tabular}
\end{center}\end{table}

\mysection{Relation to dispersive approach}
Here we study the non-perturbative contribution to the thrust
distribution from the viewpoint of the dispersive approach proposed
in ref.~[\ref{BPY}].  As discussed there and in ref.~[\ref{NaSe}],
the effects of soft gluons on event shape variables are equivalent
to a running effective coupling only after certain kinematic
approximations, which we can clarify using this approach. 

For the kinematics, we take $p_+$ and $p_-$ along the initial
quark and antiquark directions and set $2(p_+p_-)=1$ (i.e.\ we
measure all momenta in units of $Q$ in this section).
The light-cone (Sudakov) decomposition of the momentum of an
object with mass $m$ is then
\beq
 k = zp_+ + \alpha p_- +\kp\>, \qquad \alpha= \frac{\kps+m^2}{z}.
\eeq

\subsection{Exponentiation of ``massive'' soft gluons}
Each final (massless) soft parton $f$ contributes 
$\min\{\alpha_f,\,z_f\}$ to $(1\!-\!T)$. 
This statement is based on neglecting quark recoil: 
${\kps}_f$ as compared with $\alpha_f={\kps}_f/z_f$.

The first step consists of assembling final partons into subjets generated by
primary gluon radiation off the quark-antiquark line and substituting a
virtual (``massive'') gluon $i$ for each subjet.
It is implied in doing so that all the secondary partons belonging to a given
gluon subjet have the same sign of $z_f-\alpha_f$, that is, they lie in
the same hemisphere. We refer to this as the {\em moving-along} assumption.
If it is true (say, $\alpha_f<z_f$, i.e.\ right-movers), the total sum 
$\sum_{f(i)}\alpha_f=\alpha_i$ can be attributed to the primary gluon $i$.
(Notice that a massive parton contributes $\min\{\alpha_i,\,z_i\}$ to $1\!-\!T$
as well as a massless one.)
This makes internal jet structure insignificant and the problem essentially
Abelian.

Interchanging the definitions of $z$ and $\alpha$ for the {\em left-moving}\/
virtual gluons, so as to have $\min\{\alpha_i,\,z_i\}\equiv \alpha_i$,
we apply the identity
\beq\label{delta1T}
 \delta(1-T-\sum_i\alpha_i)= \int\frac{d\nu}{2\pi i}\> e^{\nu (1-T)}
\prod_i e^{-\nu \alpha_i}\>.
\eeq
This, together with the factorized matrix element for multiple soft gluon
radiation, results in a {\em factorized}\/ $i$-dependence and makes 
exponentiation straighforward:
\beq
\frac1{\sigma}\frac{d\sigma}{dT} = \int\frac{d\nu}{2\pi i}\>
\exp\left\{\nu (1-T)+ R(\nu)\right\}\>.
\eeq 
The radiator function $R(\nu)$ is obtained from the single
soft gluon emission probability 
\beq\label{gluemprob}
 \int\frac{dz}{z}\int\frac{d\alpha}{\alpha} \> \frac{C_F}{\pi} 
\int dm^2\,\left\{ -\frac1{\pi}
 \Im\left[\,\frac{\alpha_s(-m^2)}{m^2+i\epsilon}\,\right]\, \right\}\;,
\eeq
with $m^2\ge0$ the gluon invariant mass-squared.
Here for the sake of simplicity we have kept only the  
double-logarithmic part, corresponding to the main term $A(\as)$ in (\ref{lnJ}).

The crucial moving-along assumption may be correct 
for sufficiently {\em small}\/ 
gluon angles (say, $z_i>\kappa^2 \alpha_i$ with $\kappa\sim e$) 
but obviously fails for $z_i\simeq \alpha_i$ ($k_{||i}\simeq 0$,
``transversal'' gluons). 
We have to assume these do not modify the {\em nature}\/ of the leading
power correction though they may contribute to its magnitude.

Hereafter we introduce the arbitrary kinematical cut $\kappa$,
to quantify their effect. It should disappear after a full treatment
of a large-angle gluon emission with decay products falling
into opposite hemispheres.

\subsection{Running coupling in the radiator}
The exponent formally contains two pieces:
the ``real'' gluon contribution and that from a cut virtual one:
\beq\label{rvgluons}
 R(\nu) = \frac{C_F}{\pi}
\int_0^1dm^2\left\{ \alpha_s(0)\delta(m^2)+\frac{\rho(m^2)}{m^2} \right\}
\int_{m\kappa}^1\frac{dz}{z}\int_{m^2/z}^{z/\kappa^2}\frac{d\alpha}{\alpha}
 \left[\, e^{-\nu\alpha}-1\,\right]\;.
\eeq

The lower limit $\alpha>m^2/z$ comes from $\kps>0$.
The upper limit $\alpha<z/\kappa^2$ is the moving-along condition
enhanced by the factor $\kappa>1$  as discussed above,
and so is the lower limit in $z$: $z/\kappa^2>\alpha>m^2/z$.
It is the latter that brings in non-analyticity in $m^2$,
which is crucial for generating power corrections [\ref{BBB},\ref{BPY}].

To simplify the analysis we may differentiate with respect to $\nu$ 
and perform the $\alpha$ integration:
\beq
-\nu R_\nu(\nu) = \frac{C_F}{\pi}
\int_0^1dm^2\left\{ \alpha_s(0)\delta(m^2)+\frac{\rho(m^2)}{m^2} \right\}
\int_{m\kappa}^1\frac{dz}{z} \left[\, e^{-\nu m^2/z} - e^{-\nu z/\kappa^2}\,
\right]\;.
\eeq
The next step is to integrate with respect to $m^2$ by parts using 
(see ref.~[\ref{BPY}])
\beq
  dm^2\,\frac{\rho(m^2)}{m^2} = d\ae(m^2)\>.
\eeq
This gives
\beq
 (\alpha_s(0)-\ae(0))
 \int_0^1\frac{dz}{z} \left[\,1 - e^{-\nu z}\,\right]
 -\int_0^1{dm^2}\ae(m^2) \int_{m\kappa}^1 \left(-\frac{\nu}{z}\right) 
 \frac{dz}{z}e^{-\nu m^2/z}\;.
\eeq
The first term vanishes identically since
$$\eqalign{
 \alpha_s(k^2)&= -\int_0^\infty \frac{dm^2}{m^2+k^2}\rho(m^2)\>, \cr
\alpha_s(0) &= -\int_0^\infty \frac{dm^2}{m^2}\rho(m^2)=
- \int_0^\infty d\ae(m^2) = \as(\infty)-\ae(0) = \ae(0)\,.
}$$
Performing the $z$-integration we finally arrive at
\beq\label{finalcut}
-\nu R_\nu(\nu) = \frac{C_F}{\pi} \int_0^1\frac{dm^2}{m^2}\ae(m^2)
\left[\, e^{-\nu m^2}- e^{-\nu m/\kappa}\,\right].
\eeq
When evaluating (\ref{finalcut}) perturbatively, 
we do not distinguish between $\ae$ and $\as=\ae(1+\cO{\as^2})$.
This results in the above expression (\ref{lnJ}) with 
the leading term $A(\as)$ in the {\em physical scheme}\/ 
of refs.~[\ref{DKT},\ref{CMW}] for the coupling.
The next-to-leading term $B(\as)$ is also easy to reproduce by keeping
non-soft contributions in the elementary radiation probability 
(\ref{gluemprob}).

In the non-perturbative region $m^2<\mI^2$ we can expand the
exponentials to obtain the result in Eq.~(\ref{Tshift})
with the shift expressed in terms of the first non-analytic moment 
of the effective coupling ($A_1$ in the notation of ref.~[\ref{BPY}]).
The only difference from Eq.~(\ref{dlnJlin}) is an overall factor
of $1/\kappa$, indicating that the magnitude of the shift is
sensitive to the soft, large-angle region of gluon emission.

\mysection{Discussion}

We see from the above discussion that the
non-analyticity in $m^2$ in Eq.~(\ref{finalcut})
can be traced back to the very kinematical region 
$z\sim\alpha\sim m$ which does not respect the crucial moving-along
assumption. 
This does not affect the nature of the leading power correction
but it could affect its magnitude, since reducing
the contribution of this region scales down the shift $\delta T$ 
by the factor $\kappa$ 
which we have introduced to quantify sensitivity to 
large-angle gluon radiation. In the large-angle 
region one can expect an essential modification of the inclusive
spectral density $\rho$ due to the specific kinematics of the thrust:
the ``decay'' products of a timelike virtual gluon in this region
may make different contributions to the event shape, depending on
the kinematics of the decay. In the large-$N_f$ model studies of
ref.~[\ref{NaSe}], this effect was not found to be large for
the thrust, but it does depend on the shape variable involved.

In the case of other jet shape variables which have the property
of exponentiation, such as the heavy jet mass [\ref{hjm}] and jet
broadening [\ref{broad}], we expect $1/Q$ corrections to be
generated by the same mechanism.  However, we found that
for these quantities the leading non-perturbative effect
is not well represented by a simple shift in the distribution.
Furthermore, the modifications due to the large-angle region
discussed above will be different for different jet shape observables.
Thus for an extension to other related jet observables 
a quantitative analysis of the large-angle region has to be pursued. 

\section*{Acknowledgements}
We have benefited from many valuable conversations on this
topic with S.\ Catani, M.\ Dasgupta, G.\ Marchesini,
A.H.\ Mueller, P.\ Nason, G.\ Salam and M.H.\ Seymour.
Yu.L.D.\ thanks the Cavendish Laboratory and B.R.W.\ thanks
the CERN Theory Division and the St Petersburg
Nuclear Physics Institute for hospitality
while part of this work was carried out.

\newpage
\noindent{\large\bf References}

\begin{enumerate}
\item\label{renormalons}
       For reviews and classic references see:\\
       V.I. Zakharov, \np{385}{452}{92};\\ 
       A.H.\ Mueller, in {\em QCD 20 Years Later}, vol.~1
       (World Scientific, Singapore, 1993).
\item\label{BBB}
   M.\ Beneke, V.M.\ Braun and V.I.\ Zakharov, \prl{73}{3058}{94};\\
   P.\ Ball, M.\ Beneke and V.M.\ Braun, \np{452}{563}{95};\\
   M.\ Beneke and V.M.\ Braun, \np{454}{253}{95}.
\item\label{effcharge}
     G.\ Grunberg, \pl{372}{121}{96}, CPTH-PC463-0896 [hep-ph/9608375];\\ 
     N.J.\ Watson, CPT-96-P-3347 [hep-ph/9606381].
\item\label{Web94}
       B.R.\ Webber, \pl{339}{148}{94};
       see also {\em Proc.\ Summer School on Hadronic Aspects of
       Collider Physics, Zuoz, Switzerland, 1994} [hep-ph/9411384].
\item\label{DokWeb95}
       Yu.L.\ Dokshitzer and B.R.\ Webber, \pl{352}{451}{95}.
\item\label{BPY}
       Yu.L.\ Dokshitzer, G.\ Marchesini and B.R.\ Webber,
       \np{469}{93}{96}.
\item\label{KorSte}
  G.P.\ Korchemsky and G.\ Sterman, \np{437}{415}{95};
  see also {\em Proc.\ 30th Rencontres de Moriond,
  Meribel-les-Allues, France, 1995} [hep-ph/9505391].
\item\label{NaSe}
       P.\ Nason and M.H.\ Seymour, \np{454}{291}{95}.
\item\label{AZ}
    R.\ Akhoury and V.I.\ Zakharov, \pl{357}{646}{95}, \np{465}{295}{96}.
\item\label{BraBen}
     V.M.\ Braun, NORDITA-96-65P [hep-ph/9610212];\\
     M.\ Beneke,  SLAC-PUB-7277  [hep-ph/9609215].
\item\label{DKT}
     Yu.L.\ Dokshitzer, V.A.\ Khoze and S.I.\ Troyan, \pr{53}{89}{96}.
\item\label{shapexp}
  DELPHI Collaboration, P.\ Abreu et al., \zp{73}{229}{97}.
\item\label{thrust}
     S.\ Catani, L.\ Trentadue, G.\ Turnock and B.R.\ Webber,
     Phys.\ Lett.\ 263B (1991) 491.
\item\label{CTTW}
     S.\ Catani, L.\ Trentadue, G.\ Turnock and B.R. Webber,
     Nucl.\ Phys.\ B407 (1993) 3.
\item\label{tdef}
       E.\ Farhi, \prl{39}{1587}{77}.
\item\label{Jetset}
       T.\ Sj\"ostrand, \cpc{39}{347}{84};\\
       M.\ Bengtsson and T.\ Sj\"ostrand, \cpc{43}{367}{87}.
\item\label{Herwig}
  G.\ Marchesini, B.R.\ Webber, G.\ Abbiendi, I.G.\ Knowles,
  M.H.\ Seymour and L.\ Stanco, \cpc{67}{465}{92}.
\item\label{tdat1}
TASSO Collaboration, W.\ Braunschweig et al., \zp{47}{187}{90}.
\item\label{tdat2}
AMY Collaboration, Y.K.\ Li et al., \pr{41}{2675}{90}.
\item\label{tdat3}
OPAL Collaboration, P.D.\ Acton et al., \zp{59}{1}{93}.
\item\label{tdat4}
ALEPH Collaboration, R.\ Barate et al., CERN-PPE-96-186 (1996).
\item\label{tdat5}
DELPHI Collaboration, P.\ Abreu et al., CERN-PPE-96-120 (1996).
\item\label{tdat6}
SLD Collaboration, K.\ Abe et al., \pr{51}{962}{95}.
\item\label{tdat7}
OPAL Collaboration, G.\ Alexander et al., \zp{72}{191}{96}.
\item\label{tdat8}
DELPHI Collaboration, P.\ Abreu et al., CERN-PPE-96-130 (1996).
\item\label{tdat9}
OPAL Collaboration, K.\ Akerstaff et al., CERN-PPE-97-015 (1997).
\item\label{CMW}
       S.\ Catani, G.\ Marchesini and B.R.\ Webber, \np{349}{635}{91}.
\item\label{hjm}
     S.\ Catani, G.\ Turnock and B.R.\ Webber,
     Phys.\ Lett.\ 272B (1991) 368.
\item\label{broad}
       S.\ Catani, G.\ Turnock and B.R. Webber, \pl{295}{269}{92}.
\end{enumerate}
\end{document}